\documentclass[pra,aps,amsmath,amssymb,amsfonts,twocolumn]{revtex4}
\usepackage{bm,mathrsfs}

\usepackage{graphicx}
\usepackage{epsfig}
\usepackage{amsmath,bbm}
\usepackage{amsfonts,amssymb}
\usepackage{times}
\usepackage{verbatim}
\usepackage[sort&compress]{natbib}

\usepackage{amsmath}
\usepackage[colorlinks,breaklinks,linkcolor=blue,anchorcolor=blue,citecolor=blue,urlcolor=blue,
dvipdfm
]{hyperref}

\newcommand{\be}{\begin{equation}}
\newcommand{\ee}{\end{equation}}
\newcommand{\bea}{\begin{eqnarray}}
\newcommand{\eea}{\end{eqnarray}}

\def\>{\rangle}
\def\<{\langle}

\def\qed{\leavevmode\unskip\penalty9999 \hbox{}\nobreak\hfill
     \quad\hbox{\leavevmode  \hbox to.77778em{%
               \hfil\vrule   \vbox to.675em%
               {\hrule width.6em\vfil\hrule}\vrule\hfil}}
     \par\vskip3pt}
     
\begin{document}

\newtheorem{theorem}{Theorem}
\newtheorem{lemma}[theorem]{Lemma}
\newtheorem{corollary}[theorem]{Corollary}
\newtheorem{proposition}[theorem]{Proposition}
\newtheorem{definition}[theorem]{Definition}
\newtheorem{example}[theorem]{Example}
\newtheorem{conjecture}[theorem]{Conjecture}

\title{Rapid population transfer of a two-level system by a polychromatically driving field}
\author{D. X. Li}
\affiliation{Center for Quantum Sciences and School of Physics, Northeast Normal University, Changchun, 130024, China}
\affiliation{Center for Advanced Optoelectronic Functional Materials Research, and Key Laboratory for UV Light-Emitting Materials and Technology
of Ministry of Education, Northeast Normal University, Changchun 130024, China}

\author{X. Q. Shao\footnote{Corresponding author: shaoxq644@nenu.edu.cn}}
\affiliation{Center for Quantum Sciences and School of Physics, Northeast Normal University, Changchun, 130024, China}
\affiliation{Center for Advanced Optoelectronic Functional Materials Research, and Key Laboratory for UV Light-Emitting Materials and Technology
of Ministry of Education, Northeast Normal University, Changchun 130024, China}

\begin{abstract}
We propose a simple exact analytical solution for a model consisting of a two-level system and a polychromatically driving field. It helps us to realize a rapid complete population transfer from the ground state to the excited state, and the system can be stable at the excited state for an extremely long time. A combination of the mechanism and the Rydberg atoms successfully prepares the Bell state and multipartite $W$ state, and the experimental feasibility is discussed via the current experimental parameters. Finally, the simple exact analytical solution is generalized into a three-level system, which leads to a significant enhancement of the robustness against dissipation.
\end{abstract}

\maketitle



Two-level system is not only the key element in various fields of contemporary physics, such as radiation-matter interactions and collision physics \cite{pra052128ref3,pra052128ref1,pra052128ref2}, but also the fundamental building block of modern applications ranging from quantum control \cite{par023403ref4} to quantum information processing \cite{pra063414ref1.1,pra063414ref1.2}.

Moreover, the two-level system interacting with the periodically driven fields is an important prototype of a large number of quantum phenomena in nearly every subfield of optics and physics \cite{pra052128ref2}. One of the most simplest models is a  two-level system driven by a monochromatic driving field. There are also numerous interests devoting to analyzing its dynamics as the appearance of artificial two-level systems in superconducting Josephson devices \cite{pra032117ref2,pra032117ref7}, where the relevant parameters of the two-level systems can be tunable. On the other hand, a two-level system periodically driven by polychromatically driving fields results in many intriguing and important effects, \textit{e.g.} dressed-state lasers \cite{pra063809ref15.1}, multiphoton processes \cite{pra063809ref16}, polychromatically electromagnetically induced transparency \cite{pra063809ref19}, large self-phase-modulation \cite{pra063809ref20},  subhalfwavelength atom localization \cite{pra063809ref23}, resonance fluorescence \cite{PhysRevA.41.6013,PhysRevA.60.605,PhysRevA.64.013813,PhysRevA.77.063809}, decrease of ion-phonon entanglement \cite{Haddadfarshi2016} and so on.

It is well known that the typical behavior of time-dependent systems is more confusing than that of the corresponding time-independent ones. Furthermore,  the physicists' intuitions about the dynamics of time-dependent systems have not yet reached the standard of the ones for the time-independent systems. Therefore, the exact analytical solution of driven two-level problem is notoriously difficult to obtain, but it can provide us a  more transparent dynamics to  efficiently design the control strategies \cite{pra060401ref7,pra060401ref9,PhysRevLett.109.060401,PhysRevA.87.023403}. The Landau-Zener model  \cite{prl060401ref1}, the Rabi problem \cite{pra052128ref4}, and the Rosen-Zener model \cite{pra052128ref5} are the few famous examples of exactly soluble two-level evolutions. Particularly, the importance of the solution in \cite{pra052128ref5} has been demonstrated in the contexts of self-induced transparency \cite{prl060401ref6} and qubit control \cite{pra060401ref7,pra060401ref9}.  Up to now, the exploration on the exact analytical solutions still continues \cite{pra052128ref10,pra052128ref20,PhysRevA.89.043411,pra052128ref26,PhysRevA.95.052128,PhysRevA.97.022113}. Nevertheless, the known exact solutions of two-level systems are all described by the complicated functions, which include  the Gauss hypergeometric function \cite{pra052128ref10}, the Weber function \cite{prl060401ref1}, and some
more specific functions \cite{pra052128ref20,pra052128ref26}, expect for a few schemes like \cite{PhysRevA.95.052128,PhysRevA.97.022113}.

Viewpoint from the reality, it is significant to derive out the exact solutions of a two-level system driven by a polychromatically driving field, since more than one amplitude-modulated laser is applied frequently. In this letter, we exactly work out a simple analytical solution of a two-level atom interacting with a polychromatically driving field. The polychromatically driving field consists of a central field with frequency $\omega$ and $N$ pairs of symmetrically fields with frequencies $\omega\pm n\Delta$, where the central field and the symmetrically fields resonantly and dispersively drive the transitions among the states of the  two-level atom, respectively, and $n$ denotes the $n$-th pair of symmetrically detuned fields ($n=1,2,3\dots$). We find that, by means of adjusting the value of $\Delta$, a rapid complete population transfer  of the two-level atom can always occur, and then the bigger $N$ the more stable the population after the transfer. And the limiting situation, $N\rightarrow\infty$ is also discussed. In addition, we successfully achieve a seamless integration of Rydberg atoms and the polychromatically driving field to generate the Bell state and multipartite $W$ state. Finally, we generalize the model to a $\Lambda$ type atom interacting with a polychromatically driving field. Besides the analogous results to the two-level atom, we  get that, the robustness of the three-level system against atomic spontaneous emission will be remarkably improved.

Consider a two-level atom with a ground state $|g\rangle$ and an excited state $|e\rangle$, interacting with a polychromatically driving field of Rabi frequency $\Omega$. The corresponding Hamiltonian in the interaction picture can be written as
\begin{eqnarray}\label{2DfullH}
H&=&\Omega\left[1+\sum_n^N(e^{in\Delta t}+e^{-in\Delta t})\right]|e\rangle\langle g|+{\rm H.c.}\nonumber\\
&=&\Omega\left[1+2\sum_n^N\cos{(n\Delta t)}\right]|e\rangle\langle g|+{\rm H.c.}.
\end{eqnarray}
Taking advantage of the formula $\sum_n^N\cos{(n\Delta t)}=\sin{(N\Delta t+\Delta t/2)}/2\sin{(\Delta t/2)}-1/2 $, the Hamiltonian can be further simplified as
\begin{eqnarray}\label{2DH}
H&=&\frac{\sin{(N\Delta t+\Delta t/2)}}{\sin{(\Delta t/2)}}\Omega|e\rangle\langle g|+{\rm H.c.}.
\end{eqnarray}

For this system, a general wave function can be given by $|\psi(t)\rangle=c_g(t)|g\rangle+c_e(t)|e\rangle$. The equations of motion for the probability amplitudes can be obtained by the Schr\"{o}dinger equation $i|\dot\psi(t)\rangle=H|\psi(t)\rangle$ as,
$i\dot c_g(t)=A_Nc_e(t),$ and $i\dot c_e(t)=A_Nc_g(t),$
where $A_N=\Omega\sin{(N\Delta t+\Delta t/2)}/\sin{(\Delta t/2)}$. It is worthy mentioning that $A_N\rightarrow(2N+1)\Omega$ with $t\rightarrow0$. When the initial state is chosen as $|g\rangle$ ($c_g(0)=1,c_e(0)=0$), the exact analytical solutions of the probability are
\begin{eqnarray}
|c_e(t)|^2&=&\sin^2{\left[\Omega t+2\Omega\sum_n^N\frac{\sin{(n\Delta t)}}{n\Delta}\right]},\\
|c_g(t)|^2&=&\cos^2{\left[\Omega t+2\Omega\sum_n^N\frac{\sin{(n\Delta t)}}{n\Delta}\right]}.
\end{eqnarray}

While $N\rightarrow\infty$, the term $\sum_n^N\sin{(n\Delta t)/n\Delta}=i\ln{\left[-\exp{(i\Delta t)}\right]}/2\Delta$. In order to guarantee the monodromy of $\ln{\left[-\exp{(i\Delta t)}\right]}$, it can be divided into  $\ln{\left[\exp{(i(2m+1)\pi)}\right]}+ \ln{\left[\exp{(i\Delta t)}\right]}$, among which $m$ ensures that the values of $\ln{\left[\exp{(i(2m+1)\pi)}\right]}$ share the same Riemann surface with the ones of $ \ln{\left[\exp{(i\Delta t)}\right]}$ as time $t$ goes by. For instance, when time evolves into $5\pi/2\Delta$ and the principal value of $\ln{\left[\exp{(i\Delta t)}\right]}$ belongs to the range of $[2\pi,4\pi)$, $m$ needs to be $1$ to make the principal value of $\ln{\left[\exp{(i(2m+1)\pi)}\right]}$ at the same region. Therefore, when $\Delta t\in [2m'\pi,(2m'+2)\pi)$,  $m=m'~(m'=0,1,2\dots)$. Then $\ln{\left[-\exp{(i\Delta t)}\right]}=i\Delta t+i(2m+1)\pi+i2l\pi$, where $l=0,1,2\dots$ is from the periodicity of  exponential function, and its meaning is different from $m$. Hereafter, we only consider $l=0$. At last we can obtain the simplest solutions,
\begin{eqnarray}
|c_e(t)|^2&=&\sin^2{\left[\frac{(2m+1)\Omega}{\Delta}\pi\right]},\\
|c_g(t)|^2&=&\cos^2{\left[\frac{(2m+1)\Omega}{\Delta}\pi\right]}.
\end{eqnarray}
According to the simplest solutions for $N\rightarrow\infty$, we set $2\Omega/\Delta=(2j+1)/(2k+1)$, and it can be concluded that:
(i) While $j,k\in\bm {Z}$ and $(2j+1)/(2k+1)\in\bm {Z}$ ($\bm{Z}$ denotes the set of integer), the population of $|e\rangle$ will be stabilized at unity all the time, \textit{i.e.} a rapid complete population transfer occurs.
(ii) While  $j,k\in\bm {Z}$ and $(2j+1),(2k+1)$ are mutually prime, the population of $|e\rangle$ will be stabilized at unity with $\Delta t\in[(2k'k+k'-1)\pi,(2k'k+k'+1)\pi)$, $(k'=1,3,5\dots)$.
During the corresponding time, a rapid complete population transfer still occurs.
\begin{figure*}
\centering
\includegraphics[scale=0.50]{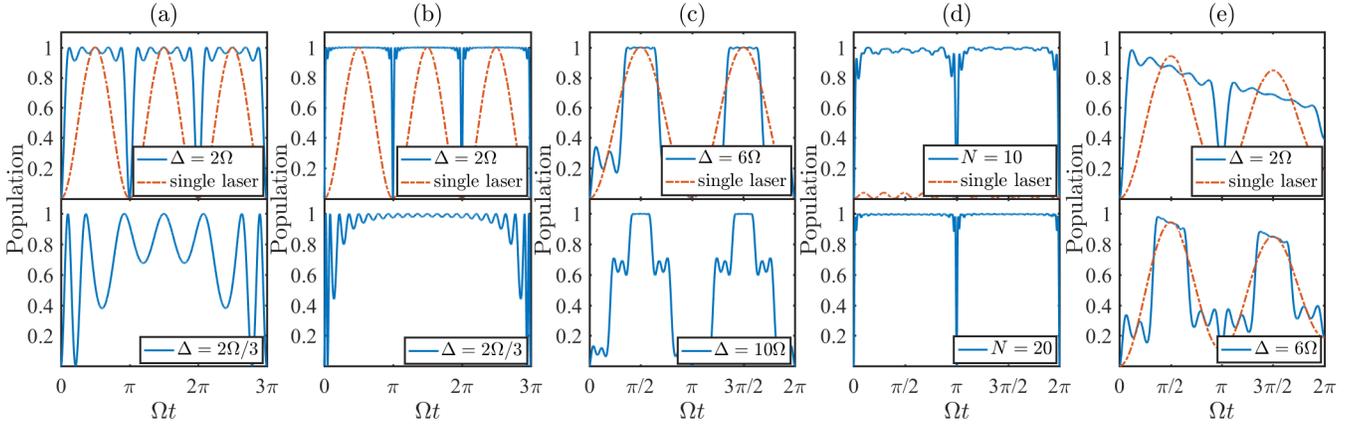}
\caption{\label{2Dzong}The populations of state $|e\rangle$ as functions of $\Omega t$ with different parameters, where the population is defined as $\langle e|\rho(t)|e\rangle$. The dashed-dotted lines indicate the situations without symmetrically detuned fields. The initial states are all the ground state. }
\end{figure*}

To demonstrate the above analyses, we plot the populations of state $|e\rangle$ as functions of $\Omega t$ with the full Hamiltonian of Eq.~(\ref{2DfullH}) governed by the Schr\"{o}dinger equation. In Fig.~\ref{2Dzong}(a) and (b), we respectively consider $N=2$ and $N=10$ to investigate the populations of state $|e\rangle$ with $\Delta=2\Omega$ and $\Delta=2\Omega/3$, where the ratio of $\Omega/\Delta$ satisfies the above conclusion (i). Compared with the situation, only one resonant central driving field (dashed-dotted line) present, the population of $|e\rangle$ with symmetrically fields possesses a higher probability  to arrive at unity. We can also find that, the more $N$, the more identical to the conclusion (i) the behaviors of state $|e\rangle$. Additionally, it is shown that for the conclusion (i), while $N$ is fixed, the effect of the rapid complete population transfer will be better with $j$ decreasing. In Fig.~\ref{2Dzong}(c) we demonstrate the conclusion (ii) with $N=2$. Although the system can't be steady at $|e\rangle$ all the time, there are still enormous advances to stabilize the system at state $|e\rangle$. Moreover, the stabilities of $|e\rangle$ in Fig.~\ref{2Dzong}(c) are superior to those in Fig.~\ref{2Dzong}(a). The former nearly exhibits a flat-top profile with $N=2$.

Distinctly, it is complicated to achieve the resonant coupling between the exited state and the ground state in experiment. Hence we suppose there is a detuning parameter $\delta$ in the process of applying the central laser to the atom. The Hamiltonian reads as
\begin{eqnarray}
H=\Omega\left[e^{i\delta t}+2\sum_n^N\cos{(n\Delta t)}\right]|e\rangle\langle g|+{\rm H.c.}.
\end{eqnarray}
In Fig.~\ref{2Dzong}(d), we study the relations of $\delta$, $N$ and the population of $|e\rangle$, where $\delta=10\Omega$ and $\Delta=2\Omega$. It is reflected that while $\delta$ is large enough to suppress the population transfer for a common two-level system (dashed-dotted line), we can introduce the symmetrically detuned fields to recover the rapid population transfer and stabilize the system at $|e\rangle$, which can be more robust against $\delta$ with bigger $N$.

In Fig.~\ref{2Dzong}(e), we take into account the atomic spontaneous emission, which can be described by Lindblad operator $L=\sqrt{\gamma}|g\rangle\langle e|$. And the corresponding master equation is
$
\dot\rho=-i[H,\rho]+L\rho L^\dag-(L^\dag L\rho+\rho L^\dag L)/2,
$
where $H$ is the full Hamiltonian of Eq.~(\ref{2DfullH}) and $\gamma=0.1\Omega,N=2$. We can learn that, despite the curve of conclusion (i) (the solid line of $\Delta=2\Omega$) more susceptible to dissipation, the population transfer of conclusion (ii) (the solid line of $\Delta=6\Omega$) is slightly better than that of only one central field (dashed-dotted line) present.

\begin{figure}[h]
\centering
\includegraphics[scale=0.54]{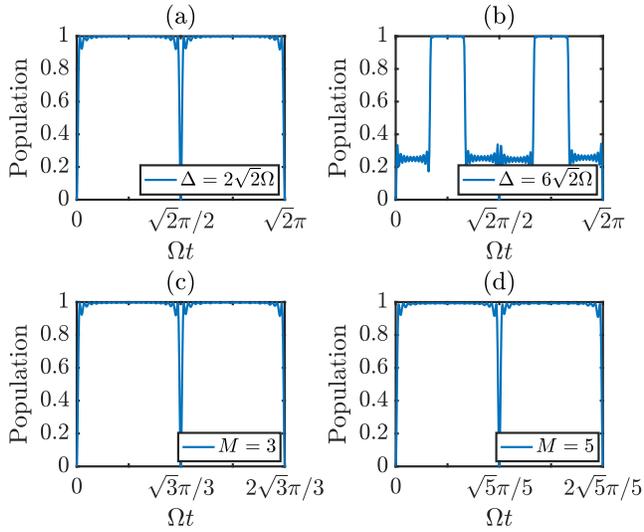}
\caption{\label{2NDentangle}(a) and (b) respectively illustrate the populations of $|T\rangle$ as functions of $\Omega t$. (c) and (d) exhibit the time evolutions of population for the multipartite $W$ state with $M=3$ and $M=5$, respectively. The other relevant parameters are: $N=10$ and $U_{\alpha\beta}=400\Omega$.}
\end{figure}
As is well known, Rydberg atoms with suitable principal quantum number can achieve long radiative lifetimes \cite{unconRPref43}, such as the 97$d_{5/2}$ Rydberg state of ${}^{87}$Rb atom with $\gamma\sim2\pi\times1$ kHz \cite{PhysRevLett.110.090402}.  Thus, we can exploit Rydberg atoms to resist detrimental effects of the atomic spontaneous emission for our scheme. Furthermore, combining  our model with the Rydberg blockade, we can prepare the Bell state $|T\rangle=(|ge\rangle+|eg\rangle)/\sqrt{2}$ and the M-qubit $W$ state $|W^M\rangle=(|g...ge)+|g...eg\rangle+\dots+|eg...g\rangle)/\sqrt{M}$, which is a crucial representative of multipartite entanglements \cite{PhysRevA.62.062314,Li:18}. The corresponding Hamiltonian is designed as
\begin{eqnarray}\label{entangle}
H&=&\sum_{\alpha=1}^M\Omega\left[1+2\sum_n^N\cos{(n\Delta t)}\right]|e\rangle_\alpha\langle g|+{\rm H.c.}\nonumber\\
&&+\sum_{\beta>\alpha}U_{\alpha\beta}|ee\rangle_{\alpha\beta}\langle ee|,
\end{eqnarray}
where $\alpha(\beta)$ stands for the $\alpha(\beta)$-th atom, and $U_{\alpha\beta}$ means the Rydberg-mediated interaction of $\alpha$-th and $\beta$-th atoms. For $M=2$ or $M>2$, it can be used to generate $|T\rangle$ or $|W^M\rangle$, respectively. Taking the case of $M=2$ as an example, the Hamiltonian can be reformulated as
\begin{eqnarray}\label{2dentangle}
H=\sqrt{2}A_N|gg\rangle\langle T|+\sqrt{2}A_N|T\rangle\langle ee|+{\rm H.c.}+U_{12}|ee\rangle\langle ee|.
\end{eqnarray}
Under the limiting condition $U_{12}\gg\sqrt{2}A_N$, the Rydberg blockade effect emerges and an effective Hamiltonian can be obtained as
\begin{eqnarray}
H_{\rm eff}=\sqrt{2}A_N|gg\rangle\langle T|+{\rm H.c.},
\end{eqnarray}
which is similar to the Eq.~(\ref{2DH}). Analogously, considering the system initialized at $|gg\rangle$, the exact analytical solution of the population for $|T\rangle$ is
\begin{eqnarray}
|c_T(t)|^2=\sin^2{\left[\sqrt{2}\Omega t+2\sqrt{2}\Omega\sum_n^N\frac{\sin{(n\Delta t)}}{n\Delta}\right]}.
\end{eqnarray}
When $N\rightarrow\infty$, we have $|c_T(t)|^2=\sin^2\left[ \sqrt{2}(2m+1)\pi\Omega/\Delta  \right]$. Set $2\sqrt{2}\Omega/\Delta=(2j+1)/(2k+1)$ and then the two conclusions will be the same as the previous (i) and (ii). As for $M>2$, by the same method, we can derive out a general effective Hamiltonian
$H^M_{\rm eff}=\sqrt{M}A_N|gg\rangle\langle W^M|+{\rm H.c.},$
general solutions
\begin{eqnarray}
|c_W^M|^2&=&\sin^2{\left[\sqrt{M}\Omega t+2\sqrt{M}\Omega\sum_n^N\frac{\sin{(n\Delta t)}}{n\Delta}\right]},\\
\lim_{N\rightarrow\infty}|c_W^M|^2&=&\sin^2\left[\frac{\sqrt{M}(2m+1)\Omega}{\Delta}\pi\right],
\end{eqnarray}
and two general conclusions after setting $2\sqrt{M}\Omega/\Delta=(2j+1)/(2k+1)$.

The $U_{\alpha\beta}$ lies on the principal quantum number, angular degrees and interatomic distance. Fortunately, our scheme just requires $U_{\alpha\beta}\gg \sqrt{M}A_N$ rather than other precisely tailored conditions. So we all assume  $U_{\alpha\beta}$ equal to $400\Omega$. In Fig.~\ref{2NDentangle}(a) and (b), the populations of $|T\rangle$ are plotted with $\Delta=2\sqrt{2}\Omega$ (satisfying conclusion (i)) and $\Delta=6\sqrt{2}\Omega$ (satisfying conclusion (ii)), respectively. The behaviors are in good agreement with those we forecast by the analytical solution. The target state can be obtained rapidly with a high population above $99\%$.
And then we research the multipartite $W$ state with $M=3,\Delta=2\sqrt{3}\Omega$ and $M=5,\Delta=2\sqrt{5}\Omega$ in Fig.~\ref{2NDentangle}(c) and (d), respectively.  The corresponding populations both arrive above $99\%$ rapidly. The validity of our scheme to generate entanglement is proven by the high degree of uniformity between the analytical and numerical results.

Then, we investigate the experimental feasibility. The Rabi laser frequency $\Omega$ can be tuned continuously between $2\pi\times (0,60)$ MHz in experiment \cite{PhysRevLett.110.090402}. After selecting the parameters as $(\Omega,\gamma)=2\pi\times(1,0.001)$ MHz, $\Delta=2\sqrt{2}\Omega$, $U_{12}=400\Omega$ and $N=4$, we calculate the population of $|T\rangle$ with master equation
$
\dot\rho=-i[H,\rho]+\sum_{\alpha=1}^2 L_\alpha\rho L_\alpha^\dag-(L_\alpha^\dag L\rho+\rho L_\alpha^\dag L_\alpha)/2,
$
 where $L_\alpha=\sqrt{\gamma}|g\rangle_\alpha\langle e|$ and $H$ is the full Hamiltonian of Eq.~(\ref{2dentangle}). The population of $|T\rangle$ will be above $98\%$ during the whole time until $t>138~\mu s$, which adequately confirms the experimental feasibility of our scheme.

\begin{figure}[h]
\centering
\includegraphics[scale=0.54]{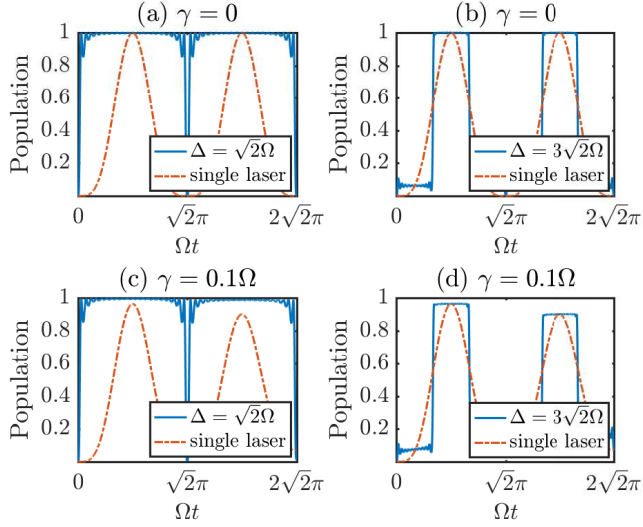}
\caption{\label{3level} The time evolutions of state $|e\rangle$ with different cases for the $\Lambda$ type atom. $N$ are all set as $10$. }
\end{figure}
More than these, the model can be also generalized to a $\Lambda$ type atom operated with a polychromatically driving field. A rapid complete population transfer from one ground state to another ground state can be realized once again, and the system will be stable at the latter. The Hamiltonian reads as
\begin{eqnarray}\label{3D}
H=A_N|e\rangle\langle r|+A_N|g\rangle\langle r|+{\rm H.c.},
\end{eqnarray}
where the states $|g\rangle$ and $|e\rangle$ are ground states, and $|r\rangle$ is the excited state. The corresponding populations can be calculated by $i\dot c_g(t)=A_Nc_r(t)$, $i\dot c_e(t)=A_Nc_r(t)$ and $i\dot c_r(t)=A_N\big[c_g(t)+c_e(t)\big]$. Substituting the initial conditions $c_g(0)=1$, $c_e(0)=0$ and $c_r(0)=0$, we can obtain the exact analytical solutions as
\begin{eqnarray}
|c_g(t)|^2&=&\cos^4\left[\sqrt{2}\Omega t/2+\sqrt{2}\Omega\sum_{n}^N\frac{\sin{(n\Delta t)}}{n\Delta}\right],\\
|c_e(t)|^2&=&\sin^4\left[\sqrt{2}\Omega t/2+\sqrt{2}\Omega\sum_{n}^N\frac{\sin{(n\Delta t)}}{n\Delta}\right],\\
|c_r(t)|^2&=&\frac{1}{2}\sin^2\left[\sqrt{2}\Omega t+2\sqrt{2}\Omega\sum_{n}^N\frac{\sin{(n\Delta t)}}{n\Delta}\right].
\end{eqnarray}
In the limit of $N\rightarrow\infty$, $|c_e(t)|^2=\sin^4\left[\sqrt{2}(2m+1)\pi\Omega/2\Delta\right]$. Set $\sqrt{2}\Omega/\Delta=(2j+1)/(2k+1)$ and the previous two conclusions are still available. The validity of conclusions (i) and (ii) is certified in Fig.~\ref{3level}(a) and (b), where the results resemble those in Fig.~\ref{2Dzong}(a)-(c).

In Fig.~\ref{3level}(c) and (d), we introduce the atomic spontaneous emission again and discuss the evolutions of $|e\rangle$ with different values of $\Delta$ respectively fulfilling conclusions (i) and (ii). The corresponding master equation can be written as
$
\dot\rho=-i[H,\rho]+\sum_{\alpha=1}^2 L_\alpha\rho L_\alpha^\dag-(L_\alpha^\dag L\rho+\rho L_\alpha^\dag L_\alpha)/2,
$
where $L_{1(2)}=\sqrt{\gamma/2}|g(e)\rangle\langle r|$ and $H$ is from Eq.~(\ref{3D}). Besides the analogous results to the two-level atom, we get that, in Fig.~\ref{3level}(c) the robustness of the three-level system against atomic spontaneous emission is remarkably improved, which even excels the situation with only one resonant central field (dashed-dotted line) present. Because the excited state $|r\rangle$ will be adiabatically eliminated as the $\Delta$ fulfilling conclusion (i), where $|c_r|^2=0$ all the while.

In summary, we have successfully derived out a simple exact analytical solution of a two-level atom interacting with a polychromatically driving field. The situations of the limiting condition are also discussed. It can guide us how to realize a rapid complete population transfer from the ground state to the excited state, and make the system stable at the excited state. Combining the analytical solutions with the Rydberg atoms, we also prepare the Bell state and the multipartite $W$ state. And the experimental feasibility is demonstrated via the state-of-the-art technology. Ultimately, the simple exact analytical solution is generalized into a $\Lambda$ type atom interacting with a polychromatically driving field. In addition to the analogous conclusions to the two-level system, we find the three-level system owns a stronger robustness against atomic spontaneous emission. we believe our work provides a new
opportunity for quantum information processing.

This work is supported by National Natural Science Foundation of China (NSFC) under Grant No. 11774047.

\bibliographystyle{apsrev4-1}
\bibliography{Poly}


\end{document}